\documentclass[10 pt]{article}
\usepackage[margin=1.5 in]{geometry}

\usepackage{array}
\usepackage{graphicx}
\usepackage{epstopdf}
\usepackage{cite}
\usepackage{amsfonts,amssymb}  
\usepackage{amsmath}
\usepackage{color}

\newtheorem{thm}{Theorem}[section]

\newtheorem{rem}[thm]{\bf Remark}

\newtheorem{clm}[thm]{\bf Claim}
\newtheorem{asm}[thm]{\bf Assumption}

\date{}

\title{\LARGE \bf  Linear Quadratic Games with Costly Measurements\footnote{This article is an unabridged version of \cite{cdc2017}.}
}

\author{Dipankar Maity\footnote{The authors are with the Department of Electrical and Computer
        Engineering and The Institute for Systems Research,
        University of Maryland, College Park, MD, USA.
        Email:
        {\tt\small dmaity@umd.edu, baras@umd.edu.}         
        Research partially supported by DARPA grant W911NF-14-1-0384 through ARO, NSF grant CNS-1544787, and DARPA STTR contract with Boston Engineering Corporation.}, Achilleas Anastasopoulos\footnote{ The author is with the Department of Electrical Engineering and Computer
        Science, University of Michigan, Ann Arbor, MI, USA. Email:
        {\tt\small anastas@umich.edu}
         }, and John S. Baras$^\dagger$ % <-this % stops a space
}

\begin{document}

\maketitle
\thispagestyle{empty}
\pagestyle{empty}

\noindent
%%%%%%%%%%%%%%%%%%%%%%%%%%%%%%%%%%%%%%%%%%%%%%%%%%%%%%%%%%%%%%%%%%%%%%%%%%%%%%%%
\begin{abstract}
In this work we consider a stochastic linear quadratic two-player game. The state
measurements are observed through a switched noiseless communication link. Each player incurs a finite
cost every time the link is established to get measurements. Along with the usual control action,
each player is equipped with a switching action to control the communication link. The measurements 
help to improve the estimate and hence reduce the quadratic cost but at the same time 
the cost is increased due to switching.
We study the subgame perfect equilibrium control and switching strategies for the players. We show that the problem can be solved in a two-step process by solving two dynamic programming problems.  The first step corresponds to solving a dynamic programming for the control strategy and the second step solves another dynamic programming for the switching strategy. 
\end{abstract}

%%%%%%%%%%%%%%%%%%%%%%%%%%%%%%%%%%%%%%%%%%%%%%%%%%%%%%%%%%%%%%%%%%%%%%%%%%%%%%%%

\section{Introduction}

Linear quadratic (LQ) stochastic games have attracted a great deal of attention in the control and related community due to its wide applicability in stochastic control, minimax control, multi-agent systems and economics \cite{basar1995dynamic},  \cite{engwerda2005lq}, \cite{foley1971class}, \cite{weeren1999asymptotic}, \cite{basar1976uniqueness},  \cite{cruz1971series},   \cite{isaacs1999differential}. There is a well established notion of (Nash) equilibrium (NE) strategies for static games, and in dynamic games there are refinements of NE known as subgame perfect equilibria (SPE). Closed form solutions for these NE (or SPE) may generally not exist or hard to compute if one such exists.
 Among the various classes of dynamic games, LQ games exhibit a closed form expression for SPE, and it is characterized by some Riccati equations. Necessary and sufficient conditions for the NE strategies of LQ games have been studied in \cite{foley1971class}, \cite{bernhard1979linear},\cite{basar1976uniqueness}. Contrary to the prior belief, \cite{basar1974counterexample} shows existence of nonlinear control strategies for LQ games. 

Amongst the vast majority of the prior works, the underlying assumption is the availability of free observations. Dynamic games are studied with either open-loop strategy (i.e. only measurement is the initial state) or feedback strategies where the observation  is available freely at any time. Challenges emerge when the measurements are on demand, but costly. This adds an extra layer of decision making, for the players, because now they have to both, control the system and ask for measurements.

\begin{figure}
\begin{center}
\includegraphics[width=0.75 \textwidth]{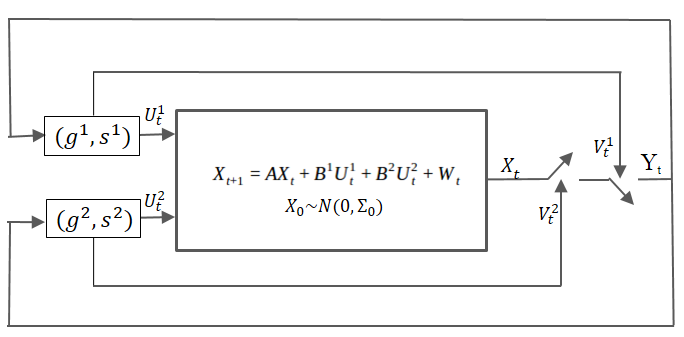}
\caption{Schematic of the system. Each player has to select their controller strategy $g^i$ and switching strategy $s^i$. All the links are noiseless and delay-free.} \label{F:schematic}
\end{center}
\end{figure}

In this work, we consider a class of two-player linear quadratic stochastic games of finite horizon. The game dynamics are partially observable. Contrary to the existing literature, the observations are not freely available. Each observation requires a finite cost for establishing a link for communication. The link through which the observations are communicated to the players (their controllers) is noiseless but operated by two switches (Figure \ref{F:schematic}), one for each player. The link is established only when both the players are willing for it, and they both get the actual state measurement at that time. Consequently, there is an apparent trade off between cost of obtaining state measurement and the estimation quality. 

In this game, the players can make a precise estimate of the state if they establish the link at every time instance. However, since the link establishment is costly, they can compromise the estimation accuracy in exchange of the cost for accessing the measurement. Therefore, the problem is to optimally decide when to establish the link and how to use the acquired measurement in order to minimize their individual cost. Since, in general, the players will have different preferences over the time instances when they want to acquire the measurement, they have to come to an agreement when to actually establish the link. 

The closest work on the similar game framework has been studied in \cite{maity2016optimal} where the authors studied zero-sum stochastic differential LQ games. However, the selection for switching times were performed in an collaborative way rather than being the outcome of a strategic interaction. The major digression of this work from \cite{maity2016optimal} is that we consider an explicit game for the switching strategy as well. We express the switch as a Boolean control action and seek for SPE for both control and switching strategies.

Our contributions are as follows:\\
(a) We study the SPE of this dynamic game and show that they can be found through a two-step process. Specifically, in the first step we fix the switching strategy and study the SPE for control strategies.
The study shows that the control strategy is linear in estimated state, where the gain is characterized with two backward Riccati equations which can be computed offline. Moreover, the Riccati equations do not depend on the switching strategy.  \\
(b) Regarding the equilibrium switching strategy, we provide a backward recursive algorithm to find all SPE where value functions need only be computed over a finite and quadratically-sized (in the duration of the game) set.\\
(c) Regarding the equilibrium switching strategy, we show that there are many equilibria among which there is one that is strictly preferable by both users and has a Markovian structure.
 It is found in our study that a strictly preferable switching strategy for a player not only depends on their own cost-to-go,  but also depends on the cost-to-go for the opponent. 

 The remaining of the paper is organized as follows: The problem formulation is provided in Section \ref{Problem Formulation}, Section \ref{Subgame Perfect Control Strategy} contains the results on the SPE of the control strategy, SPE for the switching strategy and its offline computation are analyzed in Section \ref{Subgame Perfect Switching Strategy}. Finally we conclude our work in Section \ref{Conclusion}.

\section{Problem Formulation} \label{Problem Formulation}
In the discrete time Gauss-Markov setting, we consider the following linear dynamics of the state $X_t$:

\begin{equation}
X_{t+1}=A X_t+B^1 U^1_t+B^2 U^2_t+W_t
\end{equation}
where $X_t \in \mathcal{X} = \mathbb{R}^n$, and $U^i_t \in \mathcal U^i = \mathbb{R}^m$ denotes the action of player $i$.  $W_t\in \mathbb{R}^n$ is a Gaussian noise with $\mathbb{E}[W_t]=0$  and
$\mathbb{E}[W_t W_s^{'}]=S \delta_{t-s}$ ($\delta_{t}$ is the Kronecker delta.), and $X_0 \sim \mathcal{N}(0,\Sigma_0)$.

There are two additional actions (switching actions) $V^1_t\in\{0,1\}$ and $V^2_t\in\{0,1\}$. These switching actions control a switch (switch closes if both are equal to 1) and the observation available to both users is $Y_t\in \mathcal{X} \cup \{e\}$ with
\begin{align}
Y_t = \left\{
\begin{array}{ll}
X_t &, V^1_t=V^2_t=1\\
e & , \text{else},
\end{array}
\right.
\end{align}
where ``$e$'' denotes an erasure.
The evolution of random variables in period $t$ is assumed to be
$... X_t \rightarrow (V^1_t,V^2_t) \rightarrow  Y_t \rightarrow (U^1_t,U^2_t) ...$

The information available at time $t$ to player $i$ before she takes the switching action $V^i_t$ is
\begin{align}
%I_t^i = ( Y^{t-1}, u^{i,t-1},  V^{i,t-1} ),
I_t = ( Y^{t-1}, U^{1,t-1}, U^{2,t-1}, V^{1,t-1}, V^{2,t-1} ),
\end{align}
and the information available at time $t$ to player $i$ before she takes the control action $U^i_t$ is
\begin{align}
%I_t^i = ( Y^{t-1}, u^{i,t-1},  V^{i,t-1} ),
\bar I_t = ( I_t,  V^1_t,V^2_t,Y_t ).
\end{align}
As a result, the actions have the functional form
\begin{subequations}
\begin{align}
V^i_t &= s^i_t(I_t), \qquad i=1,2, \\
U^i_t &= g^i_t(\bar I_t), \qquad i=1,2,
\end{align}
\end{subequations}
where by $g^i=(g^i_t)_{t=0}^{T-1}$, $s^i=(s^i_t)_{t=0}^{T-1}$, we denote the control and switching strategies of player $i$. $\forall t \in \{0,\cdots,T-1\}$, let us denote an $I_t$ measurable random variable $\Delta_t=V^1_t\cdot V^2_t$.

The individual cost that each player needs to minimize is quadratic in state and action, and it also depends on the switching actions $V^1_t$ and $V^2_t$.
We consider a game for a finite duration ($\{0,\cdots,T\}$) and the per-stage {costs} are explicitly written as:
\begin{align} \label{E:reward}
{C}^i_t(x_t,u^1_t,u^2_t,v^1_t,v^2_t)=&\|x_t\|^2_{Q^i}+\|u^i_t\|^2_{Q^{ii}}+\|u^j_t\|^2_{Q^{ij}}+ \lambda_i(v^i_t\cdot v^j_t) 
\end{align}
for all $t\in \{0,\cdots,T-1\}$ and
\begin{equation}
{C}_T^i(x_t,u^1_t,u^2_t,v^1_t,v^2_t)=\|x_T\|^2_{Q^i}.
\end{equation}
The quantity $\lambda_i>0$ is the cost paid by player $i$ when both the players attempt to close the switch and they observe the state information $X_t$. 
Therefore the average cost over the time horizon $\{0,\cdots,T\}$ is represented as,
\begin{equation}
J^i(\sigma^1,\sigma^2)= \sum_{t=0}^T\mathbb{E}[{C}_t^i(X_t,U^1_t,U^2_t,V^1_t,V^2_t)]
\end{equation}
where $\sigma^i=(g^i,s^i)$ denotes the strategy of the player $i$ that corresponds to control strategy $g^i$ and switching strategy $s^i$.

The objective of player $i$ is:
\begin{equation}
\min_{\sigma^i}J^i(\sigma^1,\sigma^2)=\min_{s^i}\{\min_{g^i}J^i(\sigma^1,\sigma^2)\}
\end{equation}

\section{Subgame Perfect Control Strategy} \label{Subgame Perfect Control Strategy}

For dynamic games with complete information the appropriate equilibrium concept is a refinement of Nash equilibrium (NE) called the subgame perfect equilibrium (SPE).
A strategy profile $(\sigma^1,\sigma^2)$ is a SPE if the restriction of $(\sigma^1,\sigma^2)$ to 
any proper subgame of the original game constitutes a NE \cite[pp. 94]{fudenberg1991game}.

 We seek to characterize the SPE $(\sigma^{1*},\sigma^{2*})$ for this switched LQG game.
 %\footnote{SPE are refinements of the Nash equilibria (NE) for dynamic games of complete information.}.
Moreover, we will show that among the multiple SPE, there exists one that simultaneously minimizes the cost for both users among all SPE and thus it will be the preferable SPE solution of this game.
In this section, we study the SPE control strategy for both the players.

\begin{thm} \label{T:NashControl}
\it{For any switching profile $(s^{1}, s^{2})$ of the players, the SPE control strategy $g^{i*}$ has the following structure:
\begin{align}
U^{i}_t &= g^{i*}_t(\bar I_t)= -L^i_t\hat X_t,
\end{align}
where
\begin{align} \label{E:hatx}
\hat X_{t}= \left\{
\begin{array}{ll}
A\hat X_{t-1}+B^1 U^{1}_{t-1}+B^2 U^{2}_{t-1}, & V^1_{t}\cdot V^2_{t}=0\\
X_{t},  & V^1_{t}\cdot V^2_{t}=1.
\end{array}
\right.
\end{align}

Furthermore, the cost-to-go incurred by player $i$ under the SPE control  strategy at any time step $k$ is given by,
%\begin{align} \label{E:scost}
%\mathcal{J}^i(s^1,s^2)=\mathbb{E}\Big[&\| X_{0}\|^2_{P^i_{0}}+\sum_{t=0}^{T}\big(\|E_t\|^2_{Q^i}+\\
%&(\lambda_i+\|A_{}E_{t-1}+W_{t-1}\|^2_{P^i_{t}})\cdot (V^1_t\cdot V^2_t)\big)\Big] \nonumber
%\end{align}
\begin{align} \label{E:T1}
\mathbb{J}^{i*}_k(\bar I_k)=\mathbb{E}\Big[\sum_{t=k}^{T-1}&(\|E_t\|^2_{Q^i}+\lambda_i\Delta_t)+\sum_{t=k}^{T-2}\Delta_{t+1} \|AE_{t}+W_{t}\|^2_{P^i_{t+1}}  +\|E_T\|^2_{Q^i}|~\bar I_k\Big]+\|\hat X_{k}\|_{P^i_k}^2
\end{align}
where $E_t=X_t-\hat X_t$.
The matrices $L^i_t$ and $P^i_t$ depend only on the game parameters $A,B^i,Q^i, Q^{ii}$ and $Q^{ij}$  (detailed expressions are in the proof of the theorem) and thus, can be calculated offline without the knowledge of the switching strategy profile.}
\end{thm}

\textit{proof}
The proof of this theorem is provided in Appendix \ref{A:1}.
%\end{proof}

To maintain brevity $\mathbb{J}^{i*}_k(\bar I_k)$ will be denoted as $\mathbb{J}^{i*}_k$. 
From this point onward we will set $\Delta_T=0$ and write (\ref{E:T1}) in compact form as
\begin{align} \label{E:T2}
\mathbb{J}^{i*}_k(\bar I_k)=\mathbb{E}\Big[\sum_{t=k}^{T-1}&(\|E_t\|^2_{Q^i}+\Delta_{t+1} \|AE_{t}+W_{t}\|^2_{P^i_{t+1}} +\lambda_i\Delta_t)+ \|E_T\|^2_{Q^i}|~\bar I_k\Big]+\|\hat X_{k}\|_{P^i_k}^2
\end{align}

It should be noted that in Theorem \ref{T:NashControl}, the $g^{i*}$ depends on the  given switching strategy $(s^1, s^2)$ through the $\hat X_t$. 

There are several remarks to be made at this point.

The stochastic control version of the same problem (i.e. single player single objective) is a modified Kalman filtering problem where the observations are available on demand after paying certain cost $\lambda$ per observation. Therefore, the decision of switching will solely depend on the influence of switching on the error covariance matrix. This is a side result of our work and details will appear elsewhere.

{
From Theorem \ref{T:NashControl},  $\min_{\sigma^i} J^i(\sigma^1,\sigma^2)=\min_{s^i}\mathbb{E}[\mathbb{J}^{i*}_0]$.
Therefore, the total cost incurred by player $i$ with control strategy profile ($g^{1*},g^{2*}$) is $\mathbb{E}[\mathbb{J}^{i*}_0]$. Hence, the total cost incurred with the switching is:
\begin{align} \label{AE:cost}
%&\mathcal{J}^i(s^1,s^2)=
&\mathbb{E}[\mathbb{J}^{i*}_0]=\mathbb{E}[\|\hat X_0\|_{P^i_0}^2]+\mathbb{E}\Big[\sum_{t=0}^{T-1}\big(\|E_t\|^2_{Q^i}+\Delta_{t+1} \|AE_{t}+W_{t}\|^2_{P^i_{t+1}}+\lambda_i\Delta_t\big)+\|E_T\|^2_{Q^1}\Big]. 
\end{align}
}

Another remark that is apparent from our result is that
the SPE control strategy is completely characterized by the pair of  matrices $(P^1_t, P^2_t)$  which is uniquely determined by backward dynamic equations.

%Now, we aim to find the SPE switching strategy.
%
%Defining $M_t=\mathbb E [E_t E_t']$, we have (after neglecting the term due to $X_0$)
%\begin{equation} \label{E:Js}
%\mathcal J^i(s^1,s^2)=\sum_{t=0}^{T}\mathbb E \Big[tr\big(Q^i M_t+((A_{}'P^i_{t}A_{}M_{t-1}+P^i_t
%S_{})+\lambda^i_t)\cdot(V^1_t\cdot V^2_t)\big)\Big]
%\end{equation}
%where
%\begin{align} \label{E:M}
%M_{t}&=
%\left\{
%\begin{array}{ll}
%AM_{t-1}A' +S, &  V^1_{t}\cdot V^2_{t}=0\\
%\textbf{0}, & V^1_{t}\cdot V^2_{t}=1.
%\end{array}
%\right.
%\end{align}
%Also $e_0=0\footnote{we assume the initial state is given to the players without any cost.}$ implies $M_0=\textbf{0}$ and similarly $M_{-1}=\textbf{0}$.
%
%Let us denote
%\begin{align}
%\bar C^i_t(M, 0) =&
%\begin{cases}
%tr(A'_{}Q^i_{}A_{}M+Q^i_{}S_{}) ,  & t\ne 0, \\
%0,  &t= 0 .
%\end{cases} \nonumber \\
%\bar C^i_t(M, 1) =&
%\begin{cases}
%tr(A'_{}P^i_{t}A_{}M+P^i_{t}S_{})+\lambda_i, & t \ne 0,\\
%\lambda_i, &t=0.
%\end{cases}
%\end{align}
%
%Thus,
%\begin{equation} \label{E:Js}
%\mathcal J^i(s^1,s^2)=\sum_{t=0}^{T}\mathbb E \Big[\bar C^i_t(M_{t-1},V^1_t\cdot V^2_t)\Big]
%\end{equation}
%where $M_t=\phi(M_{t-1},V^1_t\cdot V^2_t)$ is the update equation (\ref{E:M}).

\section{Subgame Perfect Switching Strategy} \label{Subgame Perfect Switching Strategy}
In this section we complete the procedure for finding the SPE of this game by focusing on the switching strategies. We will do that by considering  the backward induction process for finding SPE and reduce the cost-to-go functions into a simpler and more tractable form (compared to the one in (\ref{E:T2})).

In this problem the switching action is taken first at time $k$ based on the knowledge $I_k$ and then the augmented knowledge $\bar I_k=$ ($ I_k$, $V^1_k$, $V^2_k$, $Y_k$) is used to select the control strategies $U^i_k$. In order to visualize it, one might break the time period $[k,k+1]$ into two halves where in the first half, switching action is performed and in the second half, control action is performed. In Theorem \ref{T:NashControl}, 
$\mathbb{J}^{i*}_k$ is the optimal cost-to-go after the switching decision has been taken at time $k$.

 The actual (before switching action is taken) cost-to-go at stage $k$ is:
\begin{align}
&\mathbb{V}^{i}_k(I_k)= \mathbb{E}\Big[\sum_{t=k}^{T} C^i_t(X_t,U^1_t,U^2_t,V^1_t,V^2_t)|~ I_k\Big]
\end{align}
and the optimization (game) variables are control $U^i_t$ and switching $V^i_t$ for all $t\ge k$.

Due to the fact $\bar I_t \supseteq I_t $ for all $t$, we can write
\begin{align}
\mathbb{V}^{i}_k(I_k)= &\mathbb{E}\Big[\mathbb{E}\Big[\sum_{t=k}^{T} C^i_t(X_t,U^1_t,U^2_t,V^1_t,V^2_t)|~\bar I_k\Big]~ I_k\Big]\\
=&\mathbb{E}\big[ \mathbb{J}^i_k(g^1,g^2)|~I_k\big] \nonumber
\end{align} 
where $\mathbb{J}^i_k(g^1,g^2)=\mathbb{E}\Big[\sum_{t=k}^{T} C^i_t(X_t,U^1_t,U^2_t,V^1_t,V^2_t)|~\bar I_k\Big]$.

Since each player is interested in minimizing their cost, they are interested in $\min_{s^i,g^i}\mathbb{V}^i_k(I_k)$ at	 every stage $k$ (finally they want to minimize $\mathbb{V}^i_0(I_0)$). 

We can write,
\begin{align} \label{E:V}
\min_{s^i,g^i}\mathbb{V}^i_k(I_k)=&\min_{s^i}\{\min_{g^i}\mathbb{V}^i_k(I_k)\}=\min_{s^i}\mathbb{E}[\mathbb{J}^{i*}_k|~I_k].
\end{align}
We substitute the expression of $\mathbb{J}^{i*}_k$ from Theorem \ref{T:NashControl} into (\ref{E:V}), but before that, let us define,
 \begin{align}
 M_t =\mathbb{E}[E_tE_t'|~\bar I_t]=(1-\Delta_t)(AM_{t-1}A'+S)
 \end{align}
where  $AM_{-1}A'+S=\Sigma_0$ (since $X_0 \sim \mathcal{N}(0,\Sigma_0)$).

We also define $M_{t|t-1}=AM_{t-1}A'+S$. Note that $M_{t|t-1}$ is $I_t$ measurable whereas $M_t$ is $\bar I_t$ measurable.  $M_t$ and $M_{t|t-1}$ are related as follows:
\begin{align} \label{AE:Mfil}
M_t=(1-\Delta_t)M_{t|t-1}
\end{align}
Now let us consider the $k$-th stage cost $\mathbb{J}^{i*}_k$. 
\begin{align}
\mathbb{J}^{i*}_k=&\mathbb{E}\Big[\sum_{t=k}^{T-1}\big(\|E_t\|^2_{Q^i}+\Delta_{t+1} \|AE_{t}+W_{t}\|^2_{P^i_{t+1}} \nonumber +\lambda_i\Delta_t\big)+ \|E_T\|^2_{Q^i}|~\bar I_k\Big]+\|\hat X_{k}\|_{P^i_k}^2\nonumber \\
=&\mathbb{E}\Big[\sum_{t=k}^{T-1}\big(tr(Q^iM_{t}+\Delta_{t+1}(AM_{t}A'+S)P^i_{t+1})
+\lambda_i\Delta_t\big)+tr(Q^iM_{T})|~\bar I_k\Big] +\|\hat X_k\|_{P_k^i}^2\nonumber \\
=&\mathbb{E}\Big[\sum_{t=k}^{T-1}\big(tr((1-\Delta_t)Q^iM_{t|t-1}+ \Delta_{t+1}M_{t+1|t}P^i_{t+1})+\lambda_i\Delta_t\big)+tr(Q^iM_{T})|~\bar I_k\Big]\nonumber\\
&~~~~~~~+\|\hat X_k\|_{P_k^i}^2
%&=\sum_{t=0}^{T}\mathbb E \Big[\bar C^i_t(M_{t-1},\Delta_t)\Big]
\end{align}

Let us define $\mathcal{V}^i_k(I_k)=\mathbb E \Big[\mathbb{J}^{i*}_k|~I_k\Big]$
Therefore, 
\begin{align}
\mathcal{V}^i_k(I_k)=& \mathbb{E}\Big[\sum_{t=k}^{T-1}\big(tr((1-\Delta_t)Q^iM_{t|t-1})+tr(\Delta_{t+1}M_{t+1|t}P^i_{t+1})+\lambda_i\Delta_t\big)+tr(Q^iM_{T})|~ I_k\Big]\nonumber\\
&+\mathbb{E}\Big[\|\hat X_k\|_{P_k^i}^2|~ I_k\Big] 
\end{align}

Using (\ref{AE:filpred}), we get
\begin{align}
&\mathcal{V}^i_k(I_k)= \mathbb{E}\Big[\sum_{t=k}^{T-1}\big(tr((1-\Delta_t)Q^iM_{t|t-1})+tr(\Delta_{t}M_{t|t-1}P^i_{t})+\lambda_i\Delta_t\big)+tr(Q^iM_{T})|~ I_k\Big]\nonumber\\
&~~~~~~~~~~~~+\|\hat X_{k|k-1}\|_{P_k^i}^2 
\end{align}

The selection of switching strategy $s^i_k(I_k)$ has no effect of $\hat X_{k|k-1}$ and hence it does not play any role in the game at stage $k$.  

Let us define an instantaneous cost:

\begin{align}
\bar C^i_t(M_{t|t-1},\Delta_t)=&(1-\Delta_t)tr(Q^iM_{t|t-1})+\Delta_{t}tr(M_{t|t-1}P^i_{t})+\lambda_i\Delta_t.
\end{align}

With slight abuse of notation, after neglecting the $\hat X_{k|k-1}$ term, we obtain,
\begin{align}
&\mathcal{V}^i_k(I_k)=\mathbb{E}\Big[\sum_{t=k}^{T}\bar  C^i_t(M_{t|t-1},\Delta_t)|~I_k\Big].
\end{align}

Therefore, 
\begin{align}
\min_{s^i,g^i}\mathbb{V}^i_k(I_k)&=\min_{s^i}\mathcal{V}^i_k(I_k) \nonumber \\ &=\min_{s^i}\mathbb{E}\Big[\sum_{t=k}^{T}\bar  C^i_t(M_{t|t-1},\Delta_t)|~I_k\Big].
\end{align}

Let us denote:
\begin{equation}
\mathcal{V}^{i*}_k=\min_{s^i}\mathcal{V}^i_k(I_k).
\end{equation}

Let us perform the similar backward induction to find the SPE for the switching strategies. Note at time $T$, there is no action to optimize and 
\begin{align}
&\mathcal{V}^i_T(I_T)=\mathbb E \Big[\bar C^i_T(M_{T|T-1},\Delta_T)|~I_T\Big] =\mathbb{E}\Big[tr(Q^iM_{T})~|~I_T\Big].
\end{align}

 Let us define
\begin{align}
  \mathcal{V}^{i*}_T=\mathcal{V}^i_T(I_T)=\mathbb{E}\Big[tr(Q^iM_{T})~|~I_T\Big].
\end{align} 

Similarly, at $T-1$,
\begin{align}
\mathcal{V}^i_{T-1}(I_{T-1})=\mathbb E \Big[&\bar C^i_{T-1}(M_{T-1|T-2},\Delta_{T-1})+\bar C^i_T(M_{T},0)|~I_{T-1}\Big].
\end{align}

%$\mathcal{V}^i_{T-1}$ is the result of the SPE strategy. 

\begin{align}
\mathcal{V}^i_{T-1}(I_{T-1})=\mathbb E \Big[&\bar C^i_{T-1}(M_{T-1|T-2},\Delta_{T-1})+tr(Q^iM_{T})|~I_{T-1}\Big].
\end{align}

Using $M_{T-1}=(1-\Delta_{T-1})M_{T-1|T-2}$ and $M_T=AM_{T-1}A'+S$, 
\begin{align}
 \mathcal{V}^i_{T-1}(I_{T-1})=&\mathbb E \Big[\bar C^i_{T-1}(M_{T-1|T-2},\Delta_{T-1})+\nonumber \\&~~~(1-\Delta_{T-1})tr(Q^i(AM_{T-1|T-2}A'))+tr(Q^iS)|~I_{T-1}\Big]
\end{align}

If $(s^{1*}_{T-1}, s^{2*}_{T-1})$ is a SPE strategy at time $T-1$ then 
\begin{align}
&\mathcal{V}^i_{T-1}(I_{T-1})\big|_{(s^{i*}_{T-1}, s^{j*}_{T-1})} \le \mathcal{V}^i_{T-1}(I_{T-1})\big|_{( s^i_{T-1}, s^{j*}_{T-1})}
\end{align}
$\forall s^i_{T-1}$ and for both $i=1,2$; $j=1,2$ and $i\ne j$. 

Using the above definition of SPE, $s^i_{T-1}(I_{T-1})=0$ for $i=1,2$ is an equilibrium strategy since unilateral change from $0$ to $1$ does not change the cost for any player.  However, there might be other equilibria (in this case only $(1,1)$) which produces lower cost for the above cost function.

It is straightforward to show that the equilibrium strategy at $T-1$ is

\begin{equation} \label{E:switch}
s^{i*}_{T-1}(I_{T-1})=
\begin{cases}
1 ~~\text{ if } ~~~\bar C^i_{T-1}(M_{T-1|T-2},1) -\bar C^i_{T-1}(M_{T-1|T-2},0)\le tr(Q^i(AM_{T-1|T-2}A'))\\
0 \text{~~~otherwise}
\end{cases}
\end{equation}
%Clearly $V^i_{T-1}=1$ if $\bar C^i_{T-1}(M_{T-1|T-2},1) \le tr(Q^iS)$ and $V^i_{T-1}=0$ otherwise is an equilibrium strategy (unilateral deviation does not produce better cost).  

%{\color{blue} Due to the switching structure, cost incurred at equilibrium $(1,0)$ or ($0,1$)  is same as the cost incurred at $(0,0)$. This is because we did not have the $\epsilon_1$ cost. When $\epsilon_i$ is present, then:
% $V^i_{T-1}=1$ only if $\bar C^1_{T-1}(M_{T-2},1) \le tr(Q^1S)$ and $\bar C^2_{T-1}(M_{T-2},1) \le tr(Q^2S)$. Therefore, $V^i_{T-1}$ depends on the `optimal' cost to go for the opponent as well and wants to figure out first whether the opponent is willing to switch or not. In the case when opponent does not want to switch, then it is better to not ask for switching. 
% }

 From (\ref{E:switch}) we notice that $(1,0)$ and $(0,1)$ can also be an equilibrium strategy. However those equilibria are equivalent to $(0,0)$ in the sense that they produce the same cost-to-go $\mathcal{V}^{i*}_{T-1}$ for both $i=1,2$. Therefore, we will restrict our attention on two equilibria $(0,0)$ and $(1,1)$

As a remark, it is pointed out that
adding an infinitesimal switching cost $\epsilon_i$ for every time player $i$ requests for a switching (irrespective of whether the switch was closed or not) will ensure that $(0,1)$ and $(1,0)$ is never an SPE.

 Let us note when $\bar C^i_{T-1}(M_{T-1|T-2},1) -\bar C^i_{T-1}(M_{T-1|T-2},0)= tr(Q^i(AM_{T-1|T-2}A'))$, then $s^{i*}_{T-1}=0$ or $1$, both produces the same cost-to-go value. Under such situations, all possible switching actions are equivalent. In order to obliterate such instances we make the following assumption:
 
\begin{asm} \label{As:1}
If $\bar C^i_{T-1}(M_{T-1|T-2},1) -\bar C^i_{T-1}(M_{T-1|T-2},0)= tr(Q^i(AM_{T-1|T-2}A'))$, $s^{i*}(I_{T-1})=0$ for all possible history $I_{T-1}$. Then, (\ref{E:switch}) is modified as follows:
\begin{equation} \label{E:switch1}
s^{i*}_{T-1}(I_{T-1})=
\begin{cases}
1 ~~\text{ if } ~~~\bar C^i_{T-1}(M_{T-1|T-2},1) -\bar C^i_{T-1}(M_{T-1|T-2},0)< tr(Q^i(AM_{T-1|T-2}A'))\\
0 \text{~~~otherwise}
\end{cases}
\end{equation}
\end{asm} 
 
 Irrespective of whether SPE $s^i_{T-1}(I_{T-1})$ is $0$ or $1$, the optimal cost-to-go $\mathcal{V}^{i*}_{T-1}$ depends only on $M_{T-2}$ and also the best SPE strategy (that produces the least cost among all SPE) $s^{i*}_{T-1}(I_{T-1})$  depends only on $M_{T-2}$ (or $M_{T-1|T-2}$). 
 
Therefore, we hypothesize the following: 

\begin{clm} \label{C:markov}
For any $k$, there exists a $s^{i*}_k(I_k)$ that depends only on $M_{k-1}$ and produces the least cost-to-go among all SPE. Hence $\mathcal V^{i*}_k\equiv \mathcal V^{i*}_k(M_{k-1})$ (i.e. $\mathcal V^{i*}_k$ only depends on $M_{k-1}$). 
\end{clm}
 
 \textit{Proof:}  
 The hypothesis is true for $k=T,T-1$. Let us assume it is true for some $k+1\le T$, i.e. $\mathcal V^{i*}_{k+1}\equiv \mathcal V^{i*}_{k+1}(M_{k})$.
 Therefore,
 $$\mathcal{V}^{i*}_k=\min_{s^i}\sum_{t=k}^{T}\mathbb E \Big[\bar C^i_t(M_{t|t-1},\Delta_t)|~I_k\Big]$$
 Then using a dynamic programming argument, 
 
\begin{align} \label{AE:VJ3}
\mathcal{V}^{i*}_k=&\min_{s^i_k}\mathbb E \Big[\bar C^i_k(M_{k|k-1},\Delta_k)+\mathcal V^{i*}_{k+1}(M_k)|~I_k\Big] \nonumber \\
=&\min_{s^i_k}\mathbb E \Big[\bar C^i_k(M_{k|k-1},\Delta_k)+\mathcal V^{i*}_{k+1}\big((1-\Delta_k)M_{k|k-1}\big)|~I_k\Big]
\end{align} 

From (\ref{AE:VJ3}), the best equilibrium strategy $s^{i*}_k(I_k)=1$ if 
$$\bar C^i_k(M_{k|k-1},1)+\mathcal{V}^{i*}_{k+1}(\textbf{0}) <
\bar C^i_k(M_{k|k-1},0)+\mathcal{V}^{i*}_{k+1}(M_{k|k-1})$$  (similar to assumption \ref{As:1}, we only consider the strict inequality), otherwise $s^i_k(I_k)=0$.

Therefore $s^{i*}_k(I_k)$ requires only the knowledge of $M_{k-1}$   and hence from (\ref{AE:VJ3}), $\mathcal{V}^i_k(I_k)\equiv \mathcal{V}^{i*}_k(M_{k-1})$
% \end{proof}

 For this class of games, there always exists a  Markovian  SPE switching strategy and a Markovian SPE control strategy which produce the least cost-to-go among all SPE. Though, there might be other non-Markovian SPE strategies which produce the same cost, however, due to the claim \ref{C:markov}, it is sufficient to consider only the Markovian strategies to find the best SPE corresponding to the least cost-to-go.

\subsection{Offline Calculation of $\mathcal{V}^{i*}_k(M_{k-1})$ }

In the following we define how the players can  take the decision online by using some stored offline functions (value functions).  

%{\color{blue}
%The switching strategies can be computed offline like the gain matrices for the controller strategies. I believe, this should be the case as well... all these cost are incurred due to noise $W_t$ (its variance) and we exactly know how switching controls this variance propagation. Even though we are taking the very general case that $s^i$ is $I_t$ measurable, but intuitively thinking, the only randomness in $I_t$ is due to $W_t$ only at switching instances... and $W_t$ are i.i.d!   And these sample values of $W_t$ do not control the cost and hence to determine when to switch we do not need $I_t$ (only $M_t$ is sufficient). So far, we only thought the dependence is $I_t \rightarrow s^i_t \rightarrow I_{t+1} \rightarrow M_{t+1}\rightarrow \cdots$. In fact, this is what is happening I believe, $s_t \rightarrow M_{t+1} \xrightarrow[\text{cost}]{\text{optimize}} s_{t+1}\rightarrow$
%}

Let us define $\mathcal{V}^{i*}_k(M)$  in the following manner:
\begin{align}
\mathcal{V}^{i*}_T(M)=\bar C^i_T(M,0). ~~~~~~~~\forall M \text{ and } i=1,2
\end{align}
and
\begin{align} \label{E:opt_cost2go}
\mathcal{V}^{i*}_k(M)= \begin{cases}
\bar C^i_k(M,1)+\mathcal{V}^{i*}_{k+1}(\textbf{0})~~~~~~~~~~~ \text{ if } \vartheta(k,M) >1,\\
\bar C^i_k(M,0)+\mathcal{V}^{i*}_{k+1}(AMA'+S)~~~~~\text{ otherwise. }
\end{cases}
\end{align}
where $\vartheta(k,M)=\min\{\vartheta^1(k,M),\vartheta^2(k,M) \}$, and
\begin{align}
\vartheta^i(k,M)=\frac{\bar C^i_k(M,0)+\mathcal{V}^{i*}_{k+1}(AMA'+S)}{\bar C^i_k(M,1)+\mathcal{V}^{i*}_{k+1}(\textbf{0})}
\end{align}

%For the subgame starting at time $T$, $\mathcal{V}^{i*}_{T}(M)$ denotes the minimum cost-to-go and the equilibrium at this stage is $s^i_T(I_T)=0$ for all $I_T, M$ and $i=1,2$. 
By construction, if $\mathcal{V}^{i*}_{k+1}(\cdot)$ denotes the minimum cost-to-go (for the subgame starting at $k+1$) among the SPE, $\mathcal{V}^{i*}_{k}(\cdot)$  defined in (\ref{E:opt_cost2go}) provides the minimum cost-to-go at stage $k$ for player $i$. Therefore, by backward inductions, $\mathcal V^{i*}_\cdot(\cdot)$ denotes the cost-to-go function along an SPE that simultaneously minimizes the cost-to-go for both players.

% For a subgame starting at time $k$ and state $M$, $\mathcal{V}^{i*}(k,M)$ denotes cost-to-go for a SPE  that simultaneously minimizes the cost-to-go value for both the players among all SPE of that subgame.
%
%Now let $(\bar s^i)_k^T$ be a switching strategy such that $\bar s^i_k(I_k)=1$ for all $i=1,2$.
%
%Therefore,
%\begin{align}
%\mathcal{V}^i(k,M,(\bar s^1)_k^T,(\bar s^2)_k^T)=\bar C^i_k(M,1,1)+\mathcal{V}^i(k+1,\textbf{0},(\bar s^1)_{k+1}^T,(\bar s^2)_{k+1}^T)
%\end{align}
%
%Now, depending on $M$, there might exist a strategy profile $((\bar s^{1})_{k+1}^T,(\bar s^{2})_{k+1}^T)$ such that
%
%\begin{align}
%\bar C^i_k(M,1,1)+\mathcal{V}^i(k+1,\textbf{0},(\bar s^1)_{k+1}^T,(\bar s^2)_{k+1}^T) < \bar C^i_k(M,0,0)+\mathcal{V}^i(k+1,AMA'+S,(s^1)_{k+1}^T,(s^2)_{k+1}^T)
%\end{align}
\begin{clm}
 For any $k, M$ and history $I_k$, the best switching strategy (SPE) is given by $s^{i*}_k(I_k)=1$ for $i=1,2$ if and only if,
\begin{align} \label{E:condition}
\bar C^i_k(M,1)+\mathcal{V}^{i*}_{k+1}(\textbf{0}) < \bar C^i_k(M,0)+\mathcal{V}^{i*}_{k+1}(AMA'+S).
\end{align}
 Otherwise $s^{i*}_k(I_k)=0$ for $i=1,2$.
\end{clm}

\textit{proof}
$\Rightarrow$ is trivially true. 

$\Leftarrow$: 
 First, notice that we have established $s^i_k(I_k)=0$ is an SPE strategy for all $k,I_k$. Now let us assume that at some $k,M$, (\ref{E:condition}) holds, then if player $i$ selects a strategy such that $s^i_k(I_k)=0$, then the cost-to-go for player $i$ with any strategy profile $((s^1)_{k+1}^T,(s^2)_{k+1}^T)$ from time $k+1$ onward is
\begin{align} \label{AE:thrs}
&\bar C^i_k(M,0)+\mathcal{V}^{i}_{k+1}(AMA'+S,(s^1)_{k+1}^T,(s^2)_{k+1}^T) \nonumber\\ &\ge \bar C^i_k(M,0)+\mathcal{V}^{i*}_{k+1}(AMA'+S)\nonumber\\
&>\bar C^i_k(M,1)+\mathcal{V}^{i*}_{k+1}(\textbf{0})
\end{align}

Therefore, unilateral deviation is harmful (strictly non-profitable) for the player $i$, and that allows us to conclude $s^i_k(I_k)=1$ for $i=1,2$ is an equilibrium for $(k,M)$.  Therefore $(s^1_k(I_k),s^2_k(I_k))=(0,0)$ and $(1,1)$ both are equilibria. However, the cost-to-go by selecting $(1,1)$ is strictly lesser than selecting $(0,0)$, and this is, therefore, preferable by the players.
%\end{proof}

Note that, (\ref{E:opt_cost2go})  can be calculated and stored offline and (\ref{E:condition}) can be evaluated online using the stored values.

Equation (\ref{E:condition}) is equivalent to:
\begin{align}
\lambda_i < &\mathcal{V}^{i*}_{k+1}(AMA'+S)-\mathcal{V}^{i*}_{k+1}(\textbf{0})-tr\big( (P^i_k-Q^i)(AMA'+S)\big)
\end{align}
which shows a threshold policy for SPE switching.

We note that $M_{0|-1}=\Sigma_0$, therefore at time $0$ we only need the value $\mathcal{V}^{i*}_0(\Sigma_0)$ not the function $\mathcal{V}^{i*}_0(\cdot)$ in the entire space of symmetric positive semidefinite matrices. In order  to decide $(s^{1*}_0(I_0),s^{2*}_0(I_0))$ we need to know only four values  $\mathcal{V}^{i*}_1(\textbf{0}), \mathcal{V}^{i*}_1(A\Sigma_0A'+S)$ for $i=1,2$. Therefore, given the variance of $X_0$, we need to store only finite number of values to characterize all the value functions for a finite duration game.

\begin{clm}
The maximum number of values (value function evaluations) needed to be stored to calculate the switching strategies for entire game of duration $[0,T]$ is ${T(T+3)}$.
\end{clm}

\textit{proof}
Let at stage $k$, $M_{k|k-1}$ (or $M_{k-1}$) takes $n_k$ number of possible distinct values based on all possible previous history $I_k$. Therefore to determine the switching at time $k$, we need to make $n_k$ comparison tests (\ref{E:condition}) and for each test the $\mathcal{V}^{i*}_{k+1}(\textbf{0})$ term is common. Therefore we need to evaluate the value function only at $n_k+1$ number of points at time $k$.

For the switching pair $(1,1)$, $M_k=0$ (or $M_{k+1|k}=S$) and for any other possible switching profile at stage $k$, $M_k=M_{k|k-1}$.
Therefore at stage $k+1$, $n_{k+1}$ will be at most $n_k+1$ ( and $n_{k+1}$ possible values of $M_{k}$.) Therefore,
\begin{equation}
n_{k+1} \le n_k+1
\end{equation}
with $n_0=1$, we get $n_k\le k+1$.

\textit{Total value function evaluations to be stored} = $$2*\sum_{k=0}^{T-1}(n_{k}+1)\le {T(T+3)}.$$
The factor $2$ in above equation is due to the fact that we have to evaluate the value functions for both the players.
%\end{proof}

\begin{rem}
A switching is performed only when it strictly reduces the cost-to-go for both users. Therefore, each switching minimizes the welfare cost-to-go. However, the converse is not necessarily true i.e. a switching with a potential to reduce the welfare cost-to-go may not always be performed.
\end{rem}

\subsection{Centralized Optimization vs. Game Setup}
The problem we consider here is a game theoretic setup between two players with their own optimization criterion with two actions (control and switch). While they can select their controllers independently, however, their individual switching action does not affect the system (and cost) unless they switch synchronously. A valid question to ask is how a centralized agent would select its action strategies in order to optimize the welfare cost (i.e. the sum of two individual players' cost).

We have shown in Theorem \ref{T:NashControl} that the control strategy is totally characterized by Riccati equations for two-player setup.  Similar analysis would show that same characteristics for the control strategy are true for the centralized agent. However, it will have a single Riccati equation as opposed to two equations that we have. Similarly, the gain of the controller might change. Considering the symmetric case i.e. $B^1=B^2$, $Q^1=Q^2$, $Q^{12}=Q^{22}=Q^{11}=Q^{21}$ we can show that the control strategy for the centralized agent will be equivalent to the strategies of the two agents (i.e. $L_t= \begin{bmatrix}
L^1_t \\ L^2_t
\end{bmatrix}$), Therefore, for a fixed switching strategy, the optimal welfare cost is the same for both, the game setup and the centralized structure.
However, the centralized switching strategy will be different from game switching if $\lambda_1\ne \lambda_2$.

The above anomaly is seen since, in our model, the (selfish) players will not switch unless the switching strictly reduces their own cost, even though the switch might reduce the social welfare cost. However, the social welfare cost will always be minimized when we give the switching control to a centralized entity with the cost-to-go at stage $k$ being the social welfare $\mathcal{V}_k=\mathcal{V}^1_k+\mathcal{V}^2_k$. 
It is straightforward to notice $\mathcal V^*_k\le \mathcal{V}^{1*}_k+\mathcal{V}^{2*}_k$.

The centralized switching strategy is given by 
\begin{align} \label{AE:thrs2}
&\bar C_k(M,0)+\mathcal{V}^*_{k+1}(AMA'+S) >\bar C_k(M,1)+\mathcal{V}^*_{k+1}(\textbf{0})
\end{align}
where $\bar C_K(\cdot,\cdot)=\bar C^1_k(\cdot,\cdot)+\bar C^2_k(\cdot,\cdot)$.
An interesting study will be to characterize the social loss $l_k= \mathcal{V}^{1*}_k+\mathcal{V}^{2*}_k-\mathcal V^*_k$. 

This is also known as price of anarchy and it will be studied elsewhere. 

%\textcolor{blue}{What if  $\bar C^i_k(M,1)+\mathcal{V}^{i*}_{k+1}(\textbf{0}) = \bar C^i_k(M,0)+\mathcal{V}^{i*}_{k+1}(AMA'+S)$ at some $k,M$? Then the uniqueness of the best strategy is questionable. }

{
\section{Simulation Results}
We consider the following two-dimensional system to illustrate our analysis that has been carried out in the preceding sections.
\begin{align*}
X_{k+1} = 
\begin{bmatrix}
&0.4  & 0.8 \\ &-0.8 & 1
\end{bmatrix} X_k +  U^1_k-U^2_k+W_k
\end{align*}
where $X_k, U^1_k, U^2_k,W_k \in \mathbb{R}^2$ for all $k$. $W_k \sim \mathcal{N}(0,0.25\textbf{I})$. The observation cost parameters $\lambda_1=1$ and $\lambda_2=1.5$.

For the cost (\ref{E:reward}), the following parameters are taken: \\
$Q^1=\begin{bmatrix}
&0.3 &0 \\ &0 &0.7
\end{bmatrix}
$,
$Q^2=\begin{bmatrix}
&0.8 &0 \\ &0 &0.2
\end{bmatrix}
$,
$Q^{12}=Q^{21}=\textbf{0}$ and $Q^{11}=Q^{22}=\textbf{I}$.

One can show that $L^i_{t-1}=(-1)^{i-1}P^i_t(\textbf{I}+P^1_t+P^2_t)^{-1}A$. By denoting $P_t=\textbf{I}+P^1_t+P^2_t$, one can verify:
\begin{align*}
&P^i_t=Q^i+A'P^{-1}_{t+1}P^i_{t+1}(P^i_{t+1}+1)P^{-1}_{t+1}A\\
&P^i_T=Q^i
\end{align*}
We set the horizon of the game to be $T=15$ and assume that $X_0$ is known to the players i.e. $M_0=\textbf{0}$.

\begin{figure}[h]
\begin{center}
\includegraphics[width=0.7 \textwidth]{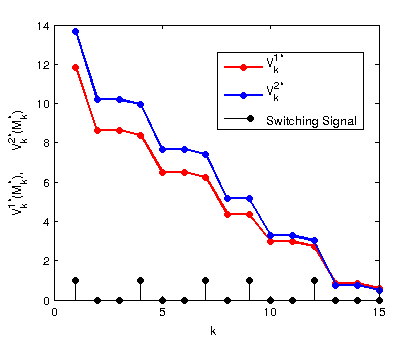}
\caption{The red and blue lines plot $\mathcal V^{i*}_k(M^*)$ w.r.t $k$ for $i=1$ and $2$ respectively. $M^*$ is the optimal trajectory of $M_k$ for the optimal switching strategy $(s^{1*},s^{2*})$. The black dots show the behavior of the optimal switching signal.} \label{F:1}
\end{center}
\end{figure}

\begin{figure}[h]
\begin{center}
\includegraphics[width=0.6 \textwidth]{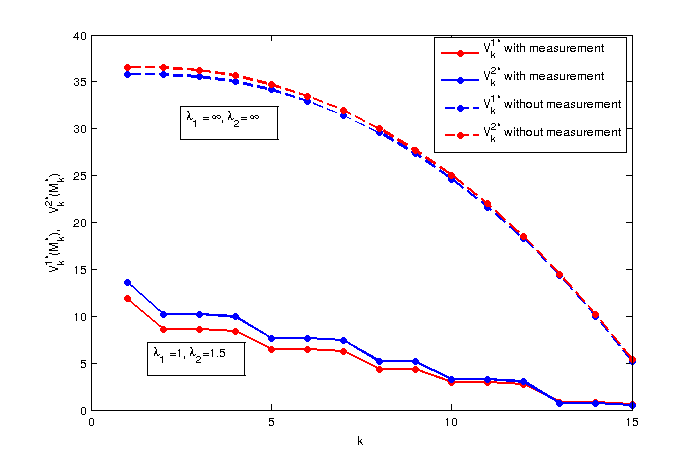}
\caption{A comparison among the costs for the cases when costly measurements are available ($\lambda_1=1, \lambda_2=1.5$) and no measurements are available ($\lambda_1=\lambda_2=\infty$)} \label{F:2}
\end{center}
\end{figure}

In Figure \ref{F:1}, we show the optimal switching strategy $\Delta_k^*(\equiv V^{1*}_k \cdot V^{2*}_k)$ in black dots. In a game with horizon 15, the switch was closed for $5$ times. In red line, we plot the value function $\mathcal V^{1*}_k(M^*_k)$ along the optimal trajectory of $M_k$ determined by (\ref{AE:Mfil}) and the optimal $\Delta_k^*$.  Similarly, in blue lines we plot $\mathcal V^{2*}_k(M_k^*)$.

In Figure \ref{F:2}, we illustrate a comparative result for the cases when observation costs are finite ($\lambda_1=1, \lambda_2=1.5$) and when observation costs are infinite (so that no observation is practically acquired). In this figure we see that even with $5$ observations (out of $15$ possible), there are more than $50\%$ reductions in costs. The dotted curves in this figure also indicate the envelop of $\mathcal V^{i*}_k$. In other words, all the graphs of $\mathcal V^{i*}_k(M^*_k)$ obtained by varying the pair $(\lambda_1, \lambda_2)$ will remain below the dotted lines shown in Figure \ref{F:2}.
}

\section{ Conclusion} \label{Conclusion}
In this work, we have considered a switched stochastic LQ game where the switching carries a finite cost. We have characterized the SPE control and switching strategies for both the players. The SPE control strategy turns out to be a linear strategy characterized by Riccati equations which do not depend on the switching strategy. The quality of state estimation depends on the switching strategy and hence the switching cost-to-go function depends on the estimation error variance. We have shown that no-switch (open switch) is always a SPE. However, at certain time instances coordinated switching is also a SPE. Moreover when both no-switch and switch are SPE, then the cost-to-go with switching is lower than the same with no-switching for both players. We studied a two-player game, however similar analysis is easily carried out for a general $n$-player game.  

\section{Appendix}

\subsection{Proof of Theorem \ref{T:NashControl}} \label{A:1}

 The idea of the proof is based on backward induction. 
 It should be noted that $\hat{X}_t$ satisfies the Kalman-filter like equations except the fact that the measurements are only available only through a switching and we always get noise free measurements whenever a switching is done.
 
We define the filtered variable as  $\hat{X}_t= \mathbb{E}[X_t~|~\bar I_t]$ and
 the prediction variable as $\hat X_{t+1|t}=\mathbb{E}[X_{t+1}|~I_t]$.

\begin{align}
\hat X_{t|t-1}=A\hat X_{t-1}+B^1 U^{1}_{t-1}+B^2 U^{2}_{t-1}
\end{align} 
 
 Therefore, $$\hat{X}_t=(1-\Delta_t)\hat X_{t|t-1}+\Delta_t X_t$$ where $\Delta_t=V^1_t\cdot V^2_t$.
 
 $\hat X_{t|t-1}$ satisfies the dynamics (\ref{E:hatx}). In a compact form, one can check
 \begin{align}
 \hat{X}_{t}=A\hat X_{t-1}+B^1 U^{1}_{t-1}+B^2 U^{2}_{t-1} +\Delta_t(AE_{t-1}+W_{t-1})
 \end{align}
 where $E_t=X_t-\hat X_t$. Thus it satisfies the difference equation:
\begin{align}\label{E:e}
E_{t}=
(1-\Delta_t)(A E_{t-1}+W_{t-1})
\end{align}

Therefore, we can write 
\begin{align} \label{AE:filpred}
\hat X_t=\hat X_{t|t-1}+\Delta_t(AE_{t-1}+W_{t-1})
\end{align}

Let $P^i_t$ satisfy the following backward equation for $i=1,2$:
\begin{align} \label{E:Pi}
P^i_t=&Q^i+L_t^i{'}Q^{ii} L^i_t+L_t^j{'}Q^{ij} L^j_t+(A-B^i L^i_t-B^j L^j_t)'P^i_{t+1}(A-B^i L^i_t-B^j L^j_t) \nonumber\\
P^i_T=&Q^i, 
\end{align}
and $L^i_t$ satisfies the relation:
\begin{align} \label{E:L}
&\Big(Q^{ii}+B^i{'}P^i_{t}\big(I-B^j(Q^{jj}+B^j{'}P^j_{t}B^j)^{-1}B^j{'}P^j_{t}\big)B^i \Big)L^i_{t-1} \nonumber \\
&=B^i{'}P^i_{t}(I-B^j(Q^{jj}+B^j{'}P^j_{t}B^j)^{-1}B^j{'}P^j_{t})A
\end{align}

%Let $t_1$ be a time such that $V^1_{t_1}=s^{1*}_{t_1}(I_{t_1})=1$ and $V^2_{t_1}=s^{2*}_{t_1}(I_{t_1})=1$. Also let $t_2>t_1$ be such that $\forall t \in [t_1+1,t_2-1]$, $V^1_{t}\cdot V^2_{t}=0$.  In other words, the given switching strategies $(s^{1*}, s^{2*})$ are such that there is no switching requested by either of the players in the time interval $[t_1+1,t_2-1]$.\\
%\\
%Thus for all $ t \in [t_1+1,t_2-1]$, $X_t=\hat X_t+E_t$.\\

Let us consider the cost segment for player $i$ for the given switching strategy profile $(s^{1*},s^{2*})$:
\begin{align} \label{E:parcost}
\mathbb{J}^{i}_k(g^1,g^2)=& \mathbb{E}\Big[\sum_{t=k}^{T} C^i_t(X_t,U^1_t,U^2_t,V^1_t,V^2_t)|~\bar I_k\Big] \nonumber \\ \nonumber
=&\mathbb{E}\Big[\sum_{t=k}^{T}C^i_t(\hat X_t,U^1_t,U^2_t,V^1_t,V^2_t)+\|E_t\|^2_{Q^i}|~\bar I_k\Big] \\ 
=&\mathbb{E}\Big[\sum_{t=k}^{T-1}(\|\hat X_t\|^2_{Q^i}+\|U^i_t\|^2_{Q^{ii}}+\|U^{j}_t\|^2_{Q^{ij}})+\| \hat X_{T}\|^2_{Q^i_{}}|~\bar I_k\Big] +\nonumber \\ &\mathbb{E}\Big[\sum_{t=k}^{T-1}(\|E_t\|^2_{Q^i}+\lambda_i\Delta_t)+\|E_T\|^2_{Q^i}|~\bar I_k\Big]
\end{align}

Let us denote $\mathbb{J}^{1*}_k=\min_{g^1} \mathbb{J}^1_k(g^1,g^{2*})$ and $g^{1*}=\arg\min_{g^1} \mathbb{J}^1_k(g^1,g^{2*})$. Similarly we define $\mathbb{J}^{2*}_k$ and $g^{2*}$. 

\begin{clm}
For all $k$, 
\begin{align}
\mathbb{J}^{i*}_k=\mathbb{E}\Big[\sum_{t=k}^{T-1}&(\|E_t\|^2_{Q^i}+\Delta_{t+1} \|AE_{t}+W_{t}\|^2_{P^i_{t+1}}
+\lambda_i\Delta_t)+\|E_T\|^2_{Q^i}|~\bar I_k\Big]+\|\hat X_{k}\|_{P^i_k}^2
\end{align}
where $\Delta_{T}=0$.
\end{clm}

\textit{proof}
This is proven by induction. It is easy to check that conditioned under $\bar I_t$, $E_t$ and $\hat X_t$ are uncorrelated.

 Hence $\mathbb{E}[\|X_t\|^2_{Q^i}|~\bar I_t]=\mathbb{E}[\|\hat X_t\|^2_{Q^i}~|~\bar I_t]+\mathbb{E}[\|E_t\|^2_{Q^i}|~\bar I_t]$.

Therefore, at $k=T$, the above claim is true. At $k=T-1$, 
\begin{align}
\mathbb{J}^{i*}_{T-1}=\min_{g^i_{T-1}}\mathbb{E}\Big[&\|\hat X_{T-1}\|^2_{Q^i}+\|U^i_{T-1}\|^2_{Q^{ii}}+\|U^{j}_{T-1}\|^2_{Q^{ij}}+ \nonumber \\ 
&\| \hat X_{T}\|^2_{Q^i_{}}|~\bar I_{T-1}\Big] +\mathbb{E}\Big[\|E_{T-1}\|^2_{Q^i}+\lambda_i\Delta_{T-1}+\|E_T\|^2_{Q^i}|~\bar I_{T-1}\Big]
\end{align}

Using $\hat X_T=A\hat X_{T-1}+B^1U^1_{T-1}+B^2U^2_{T-2}$, we obtain 
\begin{equation*}
g^{i*}_{T-1}=-L^i_{T-1}\hat X_{T-1}
\end{equation*}
Consequently, the claim holds for $k=T-1$. 

Let us assume that the claim holds for some $k+1\le T$. Then,
\begin{align} \label{AE:VJ}
\mathbb{J}^{i*}_k=&\mathbb{E}\Big[\sum_{t=k}^{T-1}(\|E_t\|^2_{Q^i}+\lambda_i\Delta_t)+\|E_T\|^2_{Q^i}|~\bar I_k\Big] + \nonumber\\ &\min_{g^{i}}\mathbb{E}\Big[\sum_{t=k}^{T-1}(\|\hat X_t\|^2_{Q^i}+\|U^i_t\|^2_{Q^{ii}}+\|U^{j}_t\|^2_{Q^{ij}})+\| \hat X_{T}\|^2_{Q^i_{}}|~\bar I_k\Big] \nonumber   \nonumber \\
=&\min_{g^i_k}\mathbb{E}\Big[ \|\hat X_k\|^2_{Q^i}+\|U^i_k\|^2_{Q^{ii}}+\|U^{j}_k\|^2_{Q^{ij}}+J^{i*}(k+1)|~\bar I_k\Big] + \mathbb{E}\Big[\|E_k\|^2_{Q^i}+\lambda_i\Delta_k|~\bar I_k\Big]
\end{align}
We used that fact that $\bar I_k \subset \bar I_{k+1}$ and hence
 $$\mathbb{E}[\mathbb{E}[X|I_{k+1}]|I_k]=\mathbb{E}[X|I_k].$$

Therefore using the hypothesis that the claim holds for $k+1$,  we can write 
 (\ref{AE:VJ}) as:
 \begin{align}
\mathbb{J}^{i*}_k =&\min_{g^i_k}\mathbb{E}\Big[ \|\hat X_k\|^2_{Q^i}+\|U^i_k\|^2_{Q^{ii}}+\|U^{j}_k\|^2_{Q^{ij}}+\|\hat X_{k+1}\|^2_{P^i_{k+1}}|~\bar I_k\Big] \nonumber \\
&+  \mathbb{E}\Big[\sum_{t=k}^{T-1}(\|E_t\|^2_{Q^i}+\lambda_i\Delta_t)+\|E_T\|^2_{Q^i}|~\bar I_k\Big]+\mathbb{E}\Big[\sum_{t=k+1}^{T-1}\Delta_{t+1} \|AE_{t}+W_{t}\|^2_{P^i_{t+1}}|~\bar I_k\Big]
\end{align}

Note that, $$\hat X_{k+1}=A\hat X_k+B^1U^1_k+B^2U^2_k +\Delta_{k+1}(AE_k+W_k)$$ and therefore, $$\mathbb{E}[\|\hat X_{k+1}\|^2_{P^i_{k+1}}|~\bar I_k]=\mathbb{E}[\|A\hat X_k+B^1U^1_k+B^2U^2_k\|^2_{P^i_{k+1}}+\Delta_{k+1}\|(AE_k+W_k)\|^2_{P^i_{k+1}}|~\bar I_k]$$. As a result, we obtain,

\begin{align} \label{AE:VJ2}
\mathbb{J}^{i*}_k
=&\min_{g^i_k}\mathbb{E}\Big[ \|\hat X_k\|^2_{Q^i}+\|U^i_k\|^2_{Q^{ii}}+\|U^{j}_k\|^2_{Q^{ij}}+  \|A\hat X_k+B^1U^1_k+B^2U^2_k\|^2_{P^i_{k+1}}|~\bar I_k\Big]  \nonumber\\
&+  \mathbb{E}\Big[\sum_{t=k}^{T-1}(\|E_t\|^2_{Q^i}+\lambda_i\Delta_t)+\|E_T\|^2_{Q^i}|~\bar I_k\Big] 
+\mathbb{E}\Big[\sum_{t=k}^{T-1}\Delta_{t+1} \|AE_{t}+W_{t}\|^2_{P^i_{t+1}}|~\bar I_k\Big]
\end{align}

Note that $\hat X_t$ is $\bar I_t$ measurable for all $t$. Thus, we can say from (\ref{AE:VJ2}) that  the optimal $U^1_{k}$ for player 1 should be given by:
\begin{align}
U^1_{k}=-(Q^{11}+B^1{'}P^1_{k+1}B^1)^{-1}B^1{'}P^1_{k+1}(A\hat X_{k}+B^2U^{2}_{k})
\end{align}
Similarly, for player $2$, it can be shown that the optimal $U^2_{k}$ will be:
\begin{align}
U^2_{k}=-(Q^{22}+B^2{'}P^2_{k+1}B^2)^{-1}B^2{'}P^2_{k+1}(A\hat X_{k}+B^1 U^{1}_{k})
\end{align}
Comparing the expressions for optimal $U^i_{k}$ and along with the definition of $L^i_k$ matrices we obtain (basically solving the two linear equations in $U^i_{k}$):
\begin{align}
U^i_{k}=g^{i*}_k(\bar I_k)=-L^i_{k}\hat X_{k}
\end{align}

Now substituting the optimal $U^i_{k}$ in (\ref{AE:VJ2}), and using the definition of $P_k^i$ from (\ref{E:Pi})  we get:
\begin{align}
\mathbb{J}^{i*}_k=\mathbb{E}\Big[\sum_{t=k}^{T-1}&(\|E_t\|^2_{Q^i}+\Delta_{t+1} \|AE_{t}+W_{t}\|^2_{P^i_{t+1}} \nonumber +\lambda_i\Delta_t)+ \|E_T\|^2_{Q^i}|~\bar I_k\Big]+\|\hat X_{k}\|_{P^i_k}^2
\end{align}
%\end{proof} 

\bibliographystyle{IEEEtran}

\bibliography{game2017}

\end{document}